\documentclass[twocolumn,english,aps,prb,superscriptaddress]{revtex4}
\usepackage[T1]{fontenc}
\usepackage{amsmath,graphicx,amssymb,xcolor}

\begin{document}
	
	\title{Long-range super-Planckian heat transfer between nanoemitters in a resonant cavity}
	\author{Kiryl Asheichyk}
	\affiliation{Department of Theoretical Physics and Astrophysics, Belarusian State University, 5 Babruiskaya Street, 220006 Minsk, Belarus}
	
	\author{Philippe Ben-Abdallah}
	\affiliation{Laboratoire Charles Fabry, UMR 8501, Institut d'Optique, CNRS, Universit\'{e} Paris-Saclay, 2 Avenue Augustin Fresnel, 91127 Palaiseau Cedex, France}
	
	\author{Matthias Kr\"{u}ger}
	\affiliation{Institute for Theoretical Physics, Georg-August-Universit\"{a}t G\"{o}ttingen, 37073 G\"{o}ttingen, Germany}
	
	\author{Riccardo Messina}
	\email{riccardo.messina@institutoptique.fr}
	\affiliation{Laboratoire Charles Fabry, UMR 8501, Institut d'Optique, CNRS, Universit\'{e} Paris-Saclay, 2 Avenue Augustin Fresnel, 91127 Palaiseau Cedex, France}
	
	\date{\today}
	
	\begin{abstract}
		We study radiative heat transfer between two nanoemitters placed inside different types of closed cavities by means of a fluctuational-electrodynamics approach. We highlight a very sharp dependence of this transfer on cavity width, and connect this to the matching between the material-induced resonance and the resonant modes of the cavity. In resonant configurations, this allows for an energy-flux amplification of several orders of magnitude with respect to the one exchanged between two emitters in vacuum as well as between two black-bodies, even at separation distances much larger than the thermal wavelength. On the other hand, variations of the cavity width by a few percent allow a reduction of the flux by several orders of magnitude and even a transition to inhibition compared to the vacuum scenario. Our results pave the way to the design of thermal waveguides for the long-distance transport of super-Planckian heat flux and selective heat transfer in many-body system.
	\end{abstract}
	
	\maketitle
	
	\section{Introduction}
	
	Two bodies kept at different temperatures and separated by vacuum exchange heat radiatively via the transfer of thermal photons. While this energy flux is limited at far separation distances by Stefan--Boltzmann's law \cite{Planck10}, the pioneering works of Rytov~\cite{Rytov53}, Polder and van Hove~\cite{Polder71} showed that this limit can be overcome in the so-called near field, i.e. when the separation distance is small compared to the thermal wavelength (around 10 microns at ambient temperature). This amplification can be remarkable for materials supporting resonant modes of the electromagnetic field~\cite{Joulain05} (in particular polar materials supporting phonon-polaritonic resonances in the infrared region of the spectrum) or a continuum of hyperbolic modes~\cite{Biehs12} (in composite materials made with a combination of dielectric and metals). Since the first theoretical investigations in the 1970s, several experiments have verified the predicted near-field heat flux amplification (see Refs.~\onlinecite{Song15,Bimonte17,Cuevas18,Biehs21} and references therein). Moreover, several applications have been put forward exploiting this strong flux increase in the near field, ranging from thermal management~\cite{Latella21a} to solid-state cooling~\cite{Chen15,Zhu19}, heat-assisted data recording and storage~\cite{Srituravanich04,BenAbdallah19,Challener09,Stipe20}, infrared sensing and spectroscopy~\cite{De Wilde06,Jones12}, energy-conversion devices~\cite{DiMatteo01,Narayanaswamy03,Laroche06,Park08,Latella21b} and thermotronics~\cite{BenAbdallah13a,BenAbdallah15}.
	
	The possibility to transport this energy from a thermal source at distances larger than its thermal wavelength remains today a challenging problem which could find broad applications in the fields of thermal management and information transfer~\cite{Rustomji19,Rustomji21,Wei21} (i.e. F\"{o}rster resonance energy transfer). Although a first strategy to achieve this transport using hyperbolic waveguides has been proposed in 2016~\cite{Messina16}, this problem remains today largely unexplored.
	
	During the last decade, a remarkable attention has been paid to the study of heat transport between two closely separated emitters or within a larger set of emitters, the main goal being to manipulate and tailor the radiative heat exchanges at the nanoscale~\cite{Messina13,BenAbdallah13b,Saaskilahti14,Biehs16,Dong18,Asheichyk18,Messina18,Deshmukh18,Tervo19,Zhang19a,Zhang19b,Luo20,Ott20,Zhang20,Ott21,Zhang21,Chen22,Fang22,Asheichyk22,Fang23,Zhang23}. Some of these studies have focused on the transfer between emitters in the proximity of a substrate ~\cite{Biehs16,Dong18,Messina18,Deshmukh18,Zhang19a,Zhang19b,Ott20,Zhang20,Ott21,Fang22} or a cylindrical waveguide~\cite{Asheichyk22}. These works discuss how the two nanoemitters can couple through the surface resonance existing at the interface between this external body and vacuum and how this additional channel contributes to the transfer by amplifying or inhibiting it. Several works have considered the impact of confinement on the heat transfer (HT) between two nanoparticles (NP), by placing them inside a planar cavity, revealing interesting effects~\cite{Saaskilahti14,Asheichyk18,Zhang21,Chen22,Zhang23}. However, they do not address at all the role played by the cavity width in the exchanged flux and especially how the presence of the cavity can make the heat transport strongly super-Planckian (far above the black-body limit) at large separation distance between the emitters.
	
	Investigating the role of confinement is the main objective of this work, where we study the near-field radiative HT between two NP placed inside a cavity (both planar and cylindrical) and highlight the crucial role played by the cavity width. More specifically, we unveil the major interplay between cavity resonant modes, whose frequencies are purely governed by the cavity width and material, and resonances existing at the interface between each NP and vacuum, giving the leading contribution to near-field radiative HT in vacuum. We show that this interplay is at the origin of a very sharp dependence of HT on cavity width, allowing for order-of-magnitude variations of HT, as well as transitions from amplification to inhibition with respect to transfer in vacuum, with variations of the cavity width in the micron range.
	
	\section{Nanoparticles inside planar and cylindrical cavities}
	
\begin{figure}[t!]
		\includegraphics[width=0.48\textwidth]{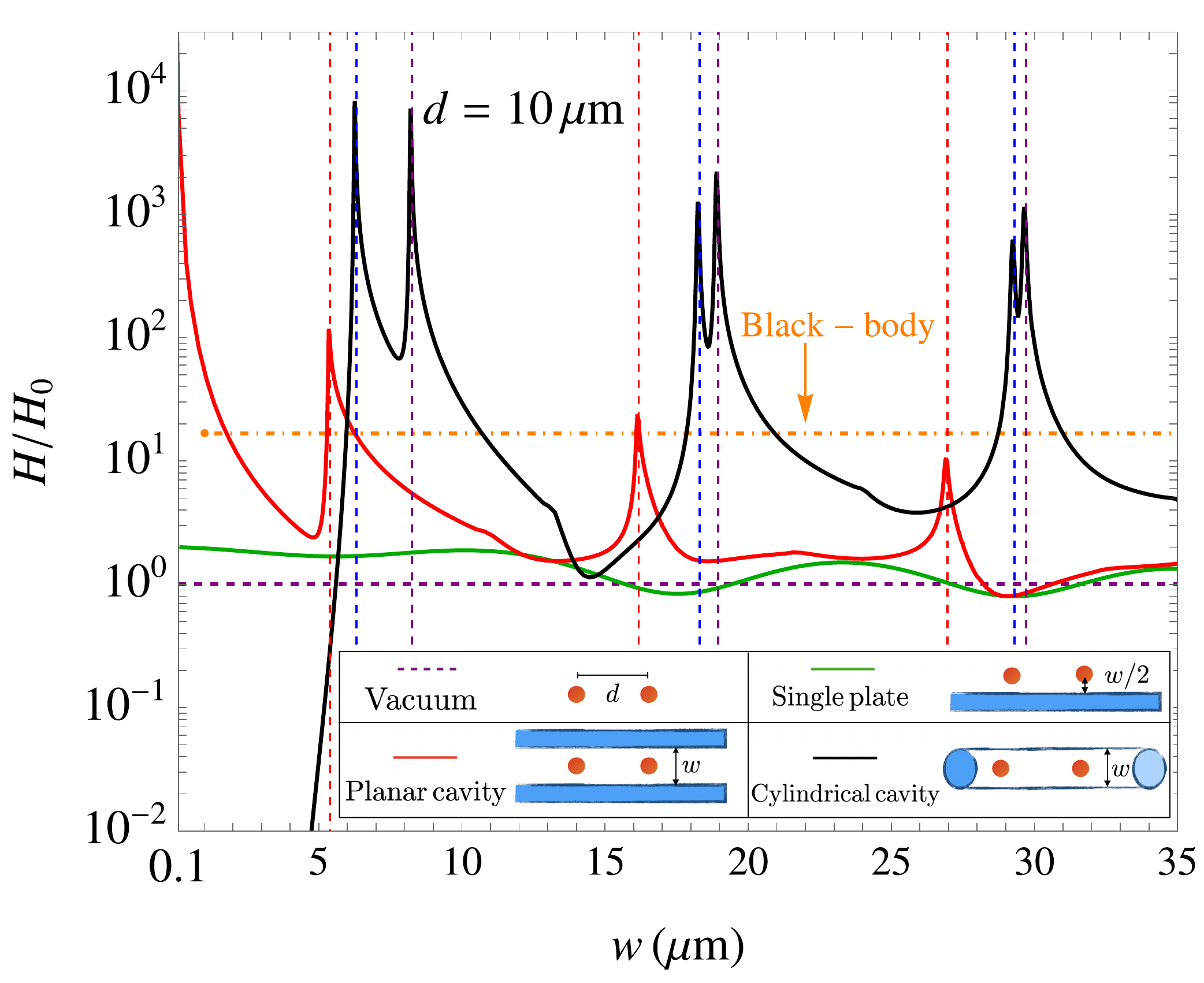}
		\caption{Ratio of heat flux $H$ in the presence of boundary conditions and the one $H_0$ in vacuum, for two NP at distance $d=10\,\mu$m. The vertical red dashed lines correspond to the planar-cavity resonances $w=(2n+1)c\pi/\omega_\mathrm{spp}$ ($n=0,1,\dots$), whereas the vertical blue and magenta dashed lines correspond to the cylindrical-cavity resonances $w=2x'_{1m}c/\omega_\mathrm{spp}$ and $w=2x_{0m}c/\omega_\mathrm{spp}$ ($m=1,2,\dots$, see text for more details), respectively. The orange dot-dashed line represents the black-body limit for two NP having a radius of 100\,nm.}
		\label{HH0_w}
	\end{figure}

	In order to highlight the effect of confinement and cavity resonances on radiative HT we consider in the following two NP placed at distance $d$, and address several different surroundings, taking as a reference the configuration in vacuum. We start with NP in proximity of a planar surface. In this configuration, as shown in \cite{Messina18}, HT can be both amplified and inhibited, depending on the distance between NP and on their distance from the substrate. We then consider two confined regimes, in which the NP are placed inside either a planar or a cylindrical cavity. The former is geometrically defined by its width $w$ (the distance between the two plates), the latter by its radius $R$\textcolor{black}{, corresponding to a cylindrical-cavity width $w=2R$}. For simplicity, the NP are always placed in the middle of the cavity (at equal distance from the two plates forming the planar cavity and along the axis of the cylindrical one). For a fair comparison, in the single-plate case the distance between NP and interface is chosen to be $w/2$. The four scenarios are depicted in the inset of Fig.~\ref{HH0_w}.
	
	The HT between the two NP in an arbitrary geometry, within the dipolar approximation, can be expressed as follows~\cite{Messina13,Asheichyk17}
	\begin{equation}\begin{split}
			H = \frac{32\pi\hbar}{c^4} \int_0^\infty d\omega\, &\omega^5[\Theta(\omega,T_1)-\Theta(\omega,T_2)]\\
			&\,\times\mathrm{Im}(\alpha_1)\mathrm{Im}(\alpha_2)\mathrm{Tr}\left(\mathbb{G}\mathbb{G}^{\dagger}\right),
			\label{eq:HT}
	\end{split}\end{equation}
	where $c$ is the speed of light, $\hbar$ and $k_{\textrm{B}}$ are Planck's and Boltzmann's constants, respectively, and $\Theta(\omega,T)=[\exp(\hbar\omega/k_{\textrm{B}}T)-1]^{-1}$, $T_1=400\,$K and $T_2=300\,$K being the temperatures of the two NP. \textcolor{black}{The temperature of the cavity walls is fixed at 300\,K, so that the colder NP exchanges heat only with the other NP}. This expression contains the imaginary part of the polarizability of two NP ($i=1,2$), for which we take the Clausius--Mossotti expression $\alpha_i (\omega)= R_i^3[\varepsilon_i(\omega)-1]/[\varepsilon_i(\omega)+2]$, where $R_i$ and $\varepsilon_i(\omega)$ are the radius and the frequency-dependent permittivity, respectively. Assuming that the two NP are made of silicon carbide (SiC), the latter is well described by the following Drude--Lorentz model~\cite{Spitzer59}
	\begin{equation}
		\varepsilon_{\textrm{SiC}}(\omega) = \varepsilon_\infty\frac{\omega^2-\omega_{\textrm{LO}}^2+i\omega\gamma}{\omega^2-\omega_{\textrm{TO}}^2+i\omega\gamma},
		\label{eq:epsilon_SiC}
	\end{equation}
	with $\varepsilon_\infty=6.7$, $\omega_{\textrm{LO}}=1.82\times10^{14}\,\textrm{rad}\,\textrm{s}^{-1}$, $\omega_{\textrm{TO}}=1.48\times10^{14}\,\textrm{rad}\,\textrm{s}^{-1}$, $\gamma=8.93\times10^{11}\,\textrm{rad}\,\textrm{s}^{-1}$. Equation \eqref{eq:HT} also contains the trace of the product of the dyadic Green's function $\mathbb{G}$, calculated at the NP coordinates, by its conjugate transpose $ \mathbb{G}^{\dagger}$. The Green's function $\mathbb{G}=\mathbb{G}_0+\mathbb{G}_T$ is written as the sum of the vacuum Green's function (known analytically~\cite{Asheichyk17}) and the scattering part $\mathbb{G}_T$, describing the effect of the boundary condition. In the following, we assume that both the planar and cylindrical boundaries are made of gold, described by the Drude model~\cite{LambrechtEPJD00} $\varepsilon_{\rm Au}(\omega) = 1-\omega_p^2/[\omega(\omega+i\omega_{\tau})]$, with $ \omega_p = 9\,{\rm eV}$ and $ \omega_{\tau} =35\,{\rm meV}$.
	
	While the expressions of $\mathbb{G}_T$ for a single plate and for a planar cavity are given in Refs.~\onlinecite{Messina18} and \onlinecite{Asheichyk18}, respectively, we present here its expression inside a cylindrical cavity of radius $R$. \textcolor{black}{Starting from the general expression of the Green's function given in Ref.~\onlinecite{GolykPRE2012} and recalling} our assumption of NP placed along the cylinder axis, $\mathbb{G}_T$ proves to be diagonal with elements 
	\begin{equation}\begin{split}
			G_{T11} = G_{T22} &= \frac{i}{8\pi}\int_0^{\infty}dk_z\Bigl[T_{1,k_z}^{MM} + 2\frac{k_z}{k}T_{1,k_z}^{MN}\\
			\,&\qquad\qquad+ \frac{k_z^2}{k^2}T_{1,k_z}^{NN}\Bigr]\cos(k_zd),\\
			G_{T33} &= \frac{i}{4\pi}\int_0^{\infty}dk_z\frac{q^2}{k^2}T_{0,k_z}^{NN}\cos(k_zd),
		\end{split}\label{eq:GT112233}\end{equation}
	expressed in terms of the scattering coefficients for \textcolor{black}{$N$ (electric) and $M$ (magnetic) polarization modes} of the field (for a non-magnetic medium). \textcolor{black}{It is easy to show that these coefficients are the same as those outside a cylinder~\cite{GolykPRE2012} of radius $R$, but with Bessel functions replaced by Hankel functions, and vice versa, as well as with an additional minus sign for $ T_{n,k_z}^{MN} $ (stemming from equality $ J_n'(qR)H_n(qR)-J_n(qR)H_n'(qR)=-2i/\pi qR$), and are given by}	
	\begin{equation}
		\begin{split}
			T_{n,k_z}^{MM} &= -\frac{H_n(qR)}{J_n(qR)}\frac{\Delta_1\Delta_4-K^2}{\Delta_1\Delta_2-K^2},\\
			T_{n,k_z}^{NN} &= -\frac{H_n(qR)}{J_n(qR)}\frac{\Delta_2\Delta_3-K^2}{\Delta_1\Delta_2-K^2},\\
			T_{n,k_z}^{MN} &= T_{n,k_z}^{NM} = -\frac{2i}{\pi\sqrt{\varepsilon}\left[qRJ_n(qR)\right]^2}\frac{K}{\Delta_1\Delta_2-K^2}.
		\end{split}\label{eq:T}\end{equation}
	Here, $J_n$ ($H_n$) is the Bessel (Hankel) function of order $n$,
	\begin{equation}
		\begin{split}
			& \Delta_1 = \frac{H'_n(q_{\varepsilon}R)}{q_{\varepsilon}RH_n(q_{\varepsilon}R)}-\frac{1}{\varepsilon}\frac{J'_n(qR)}{qRJ_n(qR)},\\
			& \Delta_2 = \frac{H'_n(q_{\varepsilon}R)}{q_{\varepsilon}RH_n(q_{\varepsilon}R)}-\frac{J'_n(qR)}{qRJ_n(qR)},\\
			& \Delta_3 = \frac{H'_n(q_{\varepsilon}R)}{q_{\varepsilon}RH_n(q_{\varepsilon}R)}-\frac{1}{\varepsilon}\frac{H'_n(qR)}{qRH_n(qR)},\\
			& \Delta_4 = \frac{H'_n(q_{\varepsilon}R)}{q_{\varepsilon}RH_n(q_{\varepsilon}R)}-\frac{H'_n(qR)}{qRH_n(qR)},
		\end{split}
	\end{equation}
	and
	\begin{equation}
		K = \frac{nk_z}{\sqrt{\varepsilon}kR^2}\left(\frac{1}{q_{\varepsilon}^2}-\frac{1}{q^2}\right),
	\end{equation}
	where $k=\omega/c$, $q = \sqrt{k^2-k_z^2}$, and $q_{\varepsilon} = \sqrt{\varepsilon k^2-k_z^2}$ the wave vector perpendicular to the axis inside the cavity walls. Note that the integrand of $G_{T11}$ ($G_{T22}$) in Eq.~\eqref{eq:GT112233} has a pole at $q = 0$ (i.e. $k_z = k$), which is integrable in principal value~\cite{Asheichyk22}, whereas the integrand of $G_{T33}$ has no poles. It can be shown that the scattering matrices, and hence the integrands of $G_{Tii}$ ($i=1,2,3$), have no poles at zeros of $J_n(qR)$ [the Bessel functions in denominators of $T_{n,k_z}^{MM}$, $T_{n,k_z}^{NN}$ and $T_{n,k_z}^{MN}$ in Eq.~\eqref{eq:T} cancel with those in $\Delta_j$, with $j=1,2,3,4$].
	
	\section{Cavity-induced amplification of heat transfer}
	
	We present in Fig.~\ref{HH0_w} the heat flux $H$ exchanged by the two SiC NP placed at distance $d=10\,\mu$m in the four scenarios described above (vacuum, single plate, planar cavity and cylindrical cavity) as a function of the cavity width $w$. In order to address directly the amplification factor, $H$ is divided by the heat flux $H_0$ exchanged in vacuum. We start by noticing that the three non-vacuum configurations result in an amplified energy flux with respect to vacuum. Nevertheless, two crucial features differentiate the single-plate configuration with respect to the two cavity (confined) ones. First, the amplification factor for a single-plate (less than 2) is strikingly over-performed by both cavities, with a flux enhancement going up to four orders of magnitude. Moreover, while the $w$ dependence in the case of a single plate is rather flat, the cavity results clearly have a resonant behavior as a function of the cavity width, suggesting the existence of a geometry-induced resonance. \textcolor{black}{We notice that the curve associated with a planar cavity has a steep behavior when the width tends to very small values. However, for such small cavity widths the validity of the dipolar approximation starts to be questioned, at least for reasonable values of the NP radius. Furthermore, the design and fabrication of such cavities is rather unpractical, and, as this amplification effect when $w$ tends to zero is not connected to the resonant behavior of interest, we do not analyze it in detail.} We also observe that, while the vacuum result is below the black-body limit (not surprisingly at $d=10\,\mu$m), shown in Fig.~\ref{HH0_w} for two NP having radii $R_1=R_2=100\,$nm, the cavity scenarios allow to produce a super-Planckian flux, with an amplification factor going up to more than two orders of magnitude for the cylindrical cavity.
	
	\begin{figure}[t!]
		\includegraphics[width=0.48\textwidth]{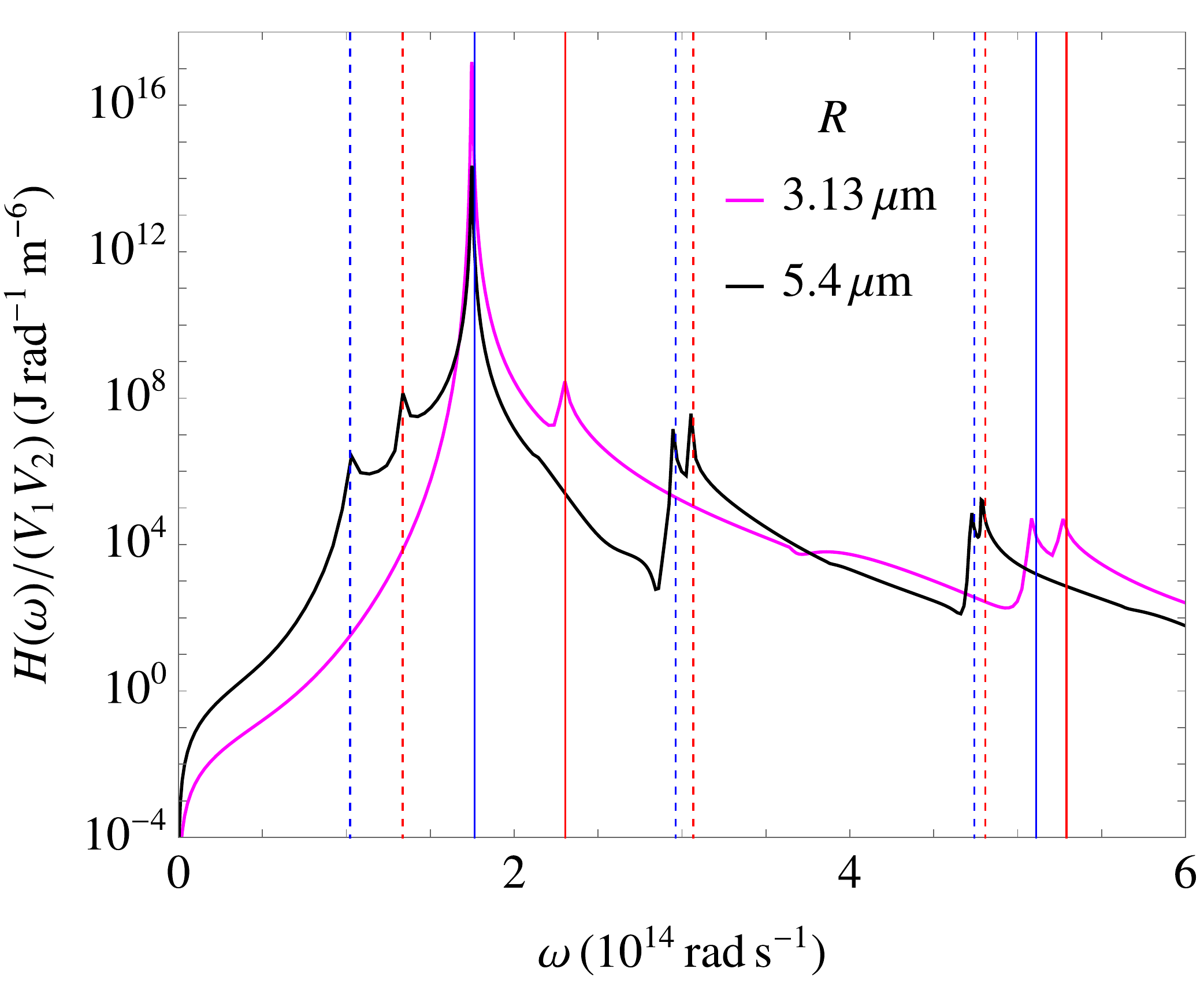}
		\caption{Spectral flux $H(\omega)$ for two NP, normalized by their volumes $V_i$, at distance $d=10\,\mu$m inside a cylindrical cavity of radius $3.13\,\mu$m (magenta) and $5.4\,\mu$m (black). The vertical blue and red lines correspond to the $R$-dependent resonances $\omega=x'_{1m}c/R$ and $\omega=x_{0m}c/R$ ($m=1,2,\dots$, see text for more details), respectively.}
		\label{H_omega}
	\end{figure}
	
	To get more insight into this resonant behavior, we start by recalling that the HT between two NP in vacuum is spectrally dominated by the surface phonon-polariton supported by each NP. This resonance occurs at the frequency $\omega_\mathrm{spp}\simeq 1.75\times10^{14}\,\mathrm{rad}\,\mathrm{s}^{-1}$ such that $\varepsilon_{\textrm{SiC}}(\omega_\mathrm{spp})\simeq-2$. We now focus on the cavities, and discuss their resonant frequencies. The simplest scenario in which these frequencies can be identified is the one when their walls are perfectly conducting. In this configuration, it is well known that imposing the boundary conditions results in an eigenvalue problem giving direct access to the cutoff frequencies~\cite{Greiner98}. In the simplest case of a planar cavity occupying the region $[-w/2,w/2]$, the $z$ dependence of the field modes has the form $\cos(k_zz)$, yielding the cutoff frequencies $\Omega_{\mathrm{pl},n}=nc\pi/w$ for $n=1,2,\dots$. Based on this analysis we may expect a matching condition of $\Omega_{\mathrm{pl},n}=\omega_\mathrm{spp}$ between cavity-induced resonance and the phonon-polariton NP resonance. Nevertheless, we should notice that for even values of $n$, the field vanishes in the center of the cavity, making these modes unable to carry energy between the two NP. We thus expect that the thicknesses $w$ resulting in a resonant amplification have the form $w=(2n+1)c\pi/\omega_\mathrm{spp}$ for $n=0,1,\dots$. This is indeed confirmed by the vertical red dashed lines in Fig.~\ref{HH0_w}. The reasoning for a cylindrical cavity is analogous, and starts from the observation [see Eq.~\eqref{eq:GT112233}], that only $M$ modes with $n=1$ and $N$ modes with $n=0,1$ can in principle contribute to the flux amplification. As discussed in \cite{Greiner98}, $N$ ($M$) modes of order $n$ have cutoff frequencies of the form $x_{nm}c/R$ ($x'_{nm}c/R$), $x_{nm}$ ($x'_{nm}$) being the $m$-th zero of $J_n(x)$ [$J'_n(x)$]. One should note that for a $N$ mode the $z$ component of the electric field is proportional to $J_n(x)$, and thus for $n=1$, being $J_1(0)=0$, the $z$ component of both electric and the magnetic fields vanish in the cavity axis. For this reason, only two sets of modes ($N$ with $n=0$ and $M$ with $n=1$) are expected to participate, giving the resonant conditions $R=x_{0m}c/\omega_\mathrm{spp}$ and $R=x'_{1m}c/\omega_\mathrm{spp}$. These resonant radii are represented by the vertical blue and magenta dashed lines in Fig.~\ref{HH0_w} and are in very good agreement with the resonances of the NP heat flux. We remark that the smallest resonant radius for a cylindrical cavity is given by $R=x'_{11}c/\omega_\mathrm{spp}\simeq 3.13\,\mu$m.
	
	While these resonances explain most of Fig.~\ref{HH0_w}, there is a difference between planar and cylindrical cavities. The curve associated with a planar cavity shows amplification when the width tends to very small values, a behavior whose investigation we leave for future work. In striking contrast, the cylindrical cavity displays pronounced inhibition for small $w$: $H$ changes by six orders of magnitude when changing $w$ by as little as two microns. \textcolor{black}{We attribute this difference to different shapes of the cavities and to different degrees of confinement (two directions are allowed for a planar cavity, whereas one direction is allowed for a cylindrical one).} This shows that such cavities modes lead to strongly selective transfer.
	
	A complementary view can be obtained by looking at the spectral heat flux $H(\omega)$ (of which the flux $H$ is the integral over frequency) for different cavity widths. The result is shown in Fig.~\ref{H_omega}, where the spectral flux associated with the first resonant radius $R=3.13\,\mu$m is compared to the one for a non-resonant radius $R=5.4\,\mu$m. We observe that both spectra peak around $\omega_\mathrm{spp}$. This is not surprising since this resonance stems from purely material properties of NP and represents the surface resonant mode existing at the interface between each NP and vacuum. The difference lies, on the contrary, in the additional cavity-induced resonances. Also in this case, their positions are very well predicted by the expressions $x_{0m}c/R$ and $x'_{1m}c/R$, which depend on the radius. This curve confirms that the strong flux amplification we obtain at $R=3.13\,\mu$m results from the matching between the first cavity-induced resonance (independent of the NP), and the one at $\omega_\mathrm{spp}$ associated with NP material properties.
	
\begin{figure}[t!]
		\includegraphics[width=0.48\textwidth]{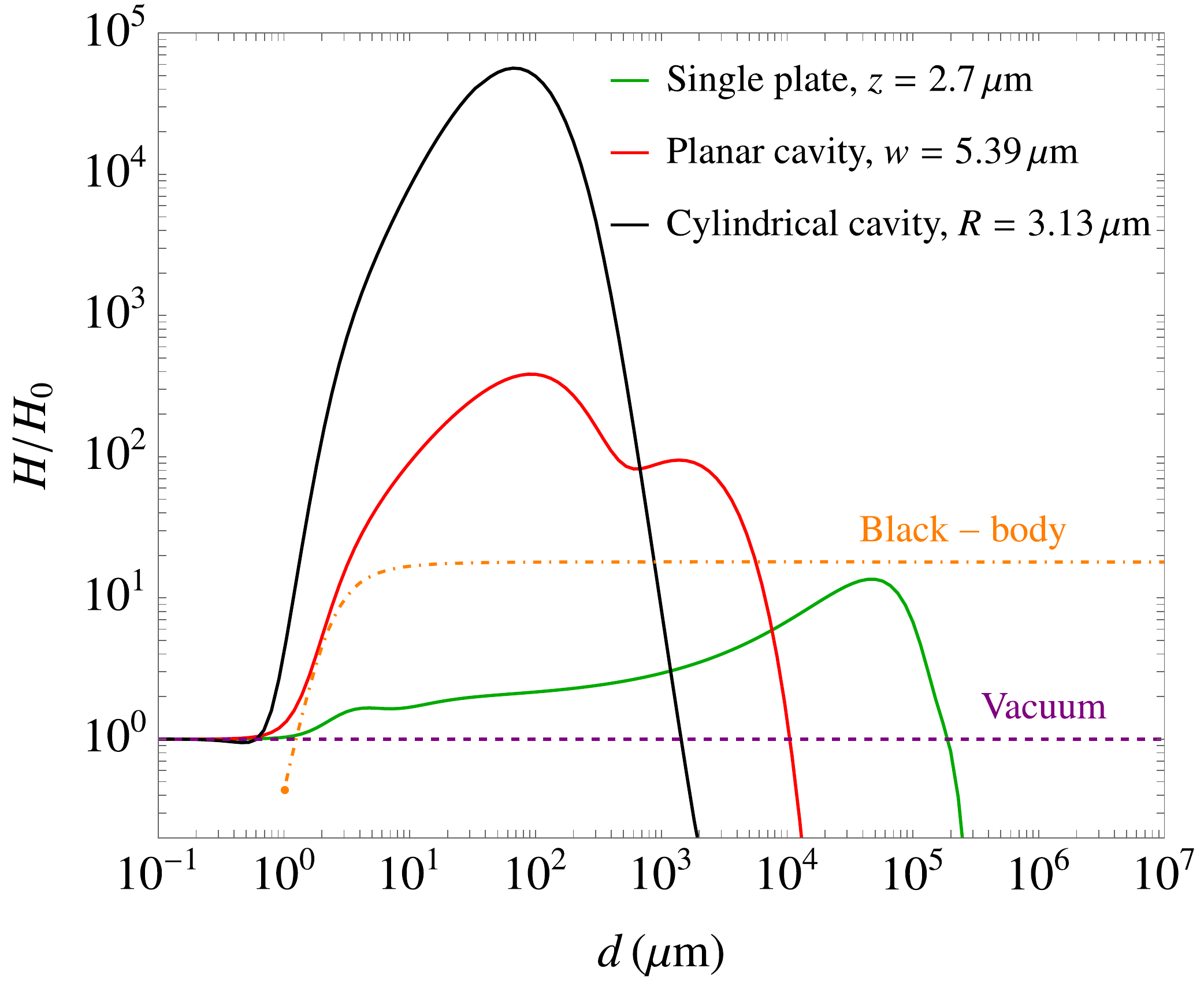}
		\caption{Ratio of heat flux $H$ in the presence of boundary conditions and the one $H_0$ in vacuum as a function of the NP distance $d$ in three different scenarios (see legend). The orange dot-dashed line represents the $d$-dependent black-body limit for two NP having a radius of 100\,nm, shown above $d=1\,\mu$m in order to be coherent with the dipolar approximation.}
		\label{HH0_d}
	\end{figure}

	As final aspect, in Fig.~\ref{HH0_d} we address the dependence of the highlighted flux amplification mechanism on the distance $d$ between the two NP. Naturally, in the graphs, we choose the optimal values of $w$ for each case, i.e, $w=5.39\,\mu$m for the planar cavity (the first resonance in Fig.~\ref{HH0_w}) and $R=w/2=3.13\,\mu$m for the cylinder. For fair comparison, the associated single-plate scenario is studied at a distance $z=2.7\,\mu$m.
	We first notice that the three curves share a common qualitative behavior, with a flux amplification starting from 1 (no amplification), increasing up to a maximum value and then decreasing and switching to a flux inhibition. The first aspect, namely the absence of amplification for small distances $d$ is expected, since in this regime, namely when $d$ is much smaller than the cavity width, the NP are strongly coupled to each other, making the presence of a boundary condition irrelevant. On the contrary, moving to larger distances makes the amplification effect manifest, and we stress that the maximum value strongly depends on the geometry under scrutiny and reaches a remarkable value of more than 4 orders of magnitude ($\simeq5\times10^4$) in the case of a cylindrical cavity, reached around $d\simeq87\,\mu$m. An interesting way of restating this result is that the cylindrical cavity is able to export to $87\,\mu$m, a distance clearly in the far-field regime, the same flux two NP would exchange in vacuum at a distance of $1.2\,\mu$m, in the near-field regime. We also notice that, while the vacuum value is below the black-body limit for distances $d\geq2\,\mu$m, the cavities allow for a super-Planckian flux with an amplification factor going beyond 3 orders of magnitude, which stays above the black-body limit up to distances of some mm. Finally, we remark that while the three geometries imply rather different maximum flux-amplification factors, the decay length of the effect they induce (the distance at which amplification turns into inhibition) also strongly depends on the geometry. Interestingly, even if for the three geometries we observe amplification up to relatively high values of the distance, this decay length actually goes in the opposite order with respect to the maximum amplification factor. In order to understand this behavior we can recall the image of heat flux between the two NP conveyed at larger distance by the surface mode existing at the cavity walls. As a consequence, we expect the decay length of amplification observed in Fig.~\ref{HH0_d} to be compatible with the one of these surface modes at the flux resonance $\omega_\mathrm{spp}$. This can be checked by seeking for the resonance (pole) of the reflection coefficient in each geometry at $\omega=\omega_\mathrm{spp}$: this gives us a parallel wavevector $k$ such that $\delta=1/\mathrm{Im}(k)$ represents an estimate of the decay length of the surface mode along the interface. This gives $\delta\simeq7\times10^4\,\mu$m for a single plate, $\delta\simeq5\times10^3\,\mu$m for a planar cavity and $\delta\simeq35\,\mu$m for a cylindrical one. These result confirm the trend observed in Fig.~\ref{HH0_d} and give a quantitative estimate of the distance at which the flux amplification reaches its maximum.
	
	\section{Conclusions}
	
	In conclusion, we have shown that confinement of two nanoemitters inside a closed cavity can induce a strong amplification or inhibition of the radiative heat flux they exchange. By analyzing two different closed geometries (a planar and a cylindrical cavity) and comparing the results to exchanges in vacuum and the simpler case of a planar interface, we have shown that this amplification stems from the matching between the spectral resonance associated with material properties (depending only on the nature of materials and on the shape of emitter) and cavity-induced resonances (depending only on the cavity geometry), and which may be used to selectively tune NP transfer. We have also shown that the characteristic distance at which the effect takes place depends on the geometry, providing a simple tool to choose the most convenient geometry depending on the desired distance. Our results pave the way to easier strategies to export and tune near-field super-Planckian radiative heat flux to distances larger than the thermal wavelength. In future work, they could be extended to other more complex geometries (curved cavity, splitter...) and the impact on many-body heat transport inside a cavity could also be addressed \cite{Zhang21,Chen22}, also addressing the role of selectivity. Finally, the role played by multipolar orders in the heat transport between nanoemitters in strongly confined systems could also be investigated.
	
	\begin{acknowledgments}
		P. B.-A. and R. M. acknowledge financial support from Labex Nanosaclay, ANR-10-LABX-0035 (Flagship project MaCaCQu). K.A. is supported by the Deutsche Forschungsgemeinschaft (DFG, German Research Foundation) through the Walter Benjamin fellowship (Project No. 453458207).
	\end{acknowledgments}

\end{document}